\documentstyle[12pt]{article}

\def\overlay#1#2{\ifmmode%
\setbox0=\hbox{$#1$}%
\setbox1=\hbox to\wd0{\hss$#2$\hss}\else%
\setbox0=\hbox{#1}%
\setbox1=\hbox to\wd0{\hss#2\hss}\fi%
 #1\hskip-\wd0\box1 }

\def\plusm#1#2{\buildrel {\raisebox{0.35ex}{\scriptsize $+#1$}} \over
{\raisebox{-0.35ex}{\scriptsize $-#2$}}}

\begin{document}
\hfill\vbox{
\hbox{FERMILAB-PUB-94-305-T}
\hbox{NUHEP-TH-94-16}
\hbox{UCD-94-28}
\hbox{hep-ph/9409316}
\hbox{August 1994} }\par
\thispagestyle{empty}

\begin{center}
{\Large \bf Perturbative QCD Fragmentation Functions as a Model for
Heavy-Quark Fragmentation}

\vspace{0.15in}

Eric Braaten\footnote{On leave from Department of Physics and Astronomy,
Northwestern University, Evanston, IL 60208} \\
{\it Theory Group, Fermilab, Batavia, IL 60510}\\

\vspace{0.15in}

Kingman Cheung\footnote{Address after Sept. 1 1994: Center for Particle
Physics, University of Texas at Austin, Austin, TX~78712} and Sean Fleming \\
{\it Department of Physics \& Astronomy, Northwestern University, Evanston,
IL 60208} \\

\vspace{0.15in}

Tzu Chiang Yuan \\

{\it Davis Institute for High Energy Physics} \\
{\it Department of Physics, University of California, Davis, CA 95616 }

\end{center}

\begin{abstract}
The perturbative QCD fragmentation functions for a heavy quark to fragment
into
heavy-light mesons are studied in the heavy-quark limit.  The
fragmentation functions for S-wave pseudoscalar and vector mesons are
calculated to
next-to-leading order in the heavy-quark mass expansion using the methods
of heavy-quark effective theory.   The results agree with the
$m_b\to \infty$ limit of the perturbative QCD fragmentation functions for
$\bar b$  into $B_c$ and $B_c^*$.
We discuss the application of the perturbative QCD fragmentation functions as
a model for the fragmentation of heavy quarks into heavy-light mesons.
Using this model, we predict the fraction $P_{V}$ of heavy-light mesons that
are produced in the vector meson state as functions of the longitudinal
momentum fraction $z$ and the transverse momentum relative to the jet axis. The
fraction $P_V$ is predicted to vary from about 1/2 at small $z$ to almost 3/4
near $z=1$.
\end{abstract}

\newpage
\begin{center}\section{Introduction}\end{center}
\label{intro}

Heavy-quark spin-flavor symmetries are very useful for understanding the
properties of hadrons containing a single heavy quark in kinematic
regimes where nonperturbative aspects of the strong interaction are dominant.
These symmetries arise from the fact that the charm, bottom, and top quarks
are much heavier than $\Lambda_{\rm QCD}$.  The symmetry is exact in the limit
of infinite quark mass, and corrections can be  systematically organized into
an expansion in powers of  $\Lambda_{\rm QCD}/m_Q$ using heavy-quark
effective theory (HQET).  There has been much progress on the applications of
heavy-quark symmetries and HQET to the spectroscopy, and to both  exclusive
and inclusive decays, of charm and bottom hadrons \cite{hq}.

It has recently been pointed out by Jaffe and Randall \cite{jaffe}
that HQET can also be  applied to the fragmentation of a heavy quark into
hadrons containing a single  heavy quark.  They showed that when the
fragmentation function is expressed in terms of an appropriate scaling
variable, it has a well-defined heavy-quark mass expansion.
Specifically, they showed that the fragmentation function
$D_{Q\to H}(z)$ at the heavy-quark mass scale has a systematic
expansion in inverse powers of $m_Q$ when expressed as a function of the
scaling variable
\begin{equation}
\label{no1}
y = \frac{1-(1-r)z}{rz} \;,
\end{equation}
where $r=(m_H - m_Q )/m_H$, $m_H$ is the mass of the heavy hadron,
and $z$ is its longitudinal momentum fraction relative to the fragmenting
heavy quark.   In the case of a heavy-light
meson, $r$ can be interpreted as the ratio of the constituent mass of the
light quark to the meson mass.
For the pseudoscalar meson $P$ and vector meson $V$ of the same S-wave
multiplet
$(^1S_0,\,^3S_1)$,   the fragmentation functions at the scale $m_Q$
have heavy-quark mass expansions of the form
\begin{eqnarray}
\label{dz1}
D_{Q\to P}(z)   &=& \frac{a(y)}{r} + b(y) + {\cal O}(r) \,, \\
\label{dz2}
D_{Q\to V}(z) &=& \frac{a^*(y)}{r} + b^*(y) + {\cal O}(r) \,,
\end{eqnarray}
where $a^*(y)=3a(y)$.
By heavy-quark spin symmetry, the leading terms differ by a spin factor of 3,
while spin splittings first appear at next-to-leading order in the functions
$b(y)$ and $b^*(y)$.

It was also realized recently that the fragmentation functions for mesons
containing a heavy quark and a heavy antiquark can be computed using
perturbative quantum chromodynamics (PQCD) \cite{gluon,chang,psi}.  The
fragmentation functions for a $\bar b$ to split into the S-wave $\bar b c$
mesons $B_c$ and $B_c^*$ were calculated to leading order in $\alpha_s$ in
Ref.~\cite{bc}.  These fragmentation functions have been used to predict
the production rates of the $B_c$ meson at LEP and at the Tevatron
\cite{bc-direct,bc-induce}.
The general analysis of Jaffe and Randall must certainly apply to perturbative
QCD fragmentation functions in the limit where the mass of the heavier quark
is taken to infinity.   It was verified explicitly in Ref.~\cite{bc}
that the PQCD fragmentation functions $D_{\bar b\to B_c}(z)$  and
$D_{\bar b\to B_c^*}(z)$ reduce to the forms (\ref{dz1}) and (\ref{dz2})
with $r=m_c/(m_b+m_c)$ in the limit $m_b\to \infty$.

Since the PQCD fragmentation functions are consistent with heavy-quark
symmetry, they can be used as models for the fragmentation of heavy quarks into
heavy-light mesons.  In this paper, we show how the leading and
next-to-leading terms in the $1/m_Q$ expansions can be calculated directly
{}from HQET.  We then discuss the use of the PQCD fragmentation functions as a
phenomenological model for the  fragmentation of charm and bottom quarks into
heavy-light mesons.  As an application of this model, we consider the fraction
$P_V$  of heavy-light mesons that are produced in the vector-meson state.

\begin{center}\section{PQCD Fragmentation Functions from HQET}\end{center}

The HQET Lagrangian, including the leading and the $1/m_Q$ terms, is given by
\cite{hq}
\begin{equation}
{\cal L} = \bar h_v \left\{ iv\cdot D
 +  \frac{1}{2m_Q} \left(
C_1(iD)^2 - C_2(v\cdot iD)^2 - \frac{C_3}{2}g_s \sigma^{\mu\nu}G_{\mu\nu}
\right ) \right \} h_v
\end{equation}
where
\begin{equation}
\label{c1-c3}
\begin{array}{rcl}
C_1 &=& 1\,, \\
C_2 &=& 3\left(\frac{\alpha_s(\mu)}{\alpha_s(m_Q)}\right)^{-8/(33-2n_f)}-2
\,,\\
C_3&=&\left( \frac{\alpha_s(\mu)}{\alpha_s(m_Q)} \right)^{-9/(33-2n_f)}\,.
\end{array}
\end{equation}
These coefficients are all equal to 1 at the heavy-quark mass scale $\mu=m_Q$.
The term proportional to $C_2$ can be  omitted in calculating physical
quantities, because it can be eliminated using a field redefinition involving
 the equation of motion $(v\cdot D)h_v=0$ from the leading term in the
Lagrangian.  Our method for calculating the fragmentation function involves
a heavy  quark which is off-shell by an amount at least of order $m_Q m_q$.
To demonstrate that the $C_2$ term can still be omitted in this case,
we keep it in our calculation throughout, and show that it
cancels between the vertex and propagator corrections.

In our calculation, we need the Feynman rules derived from the HQET Lagrangian
for (i) the heavy-quark propagator, including $1/m_Q$ corrections,
(ii) the heavy-quark-gluon vertex, including $1/m_Q$ corrections, and (iii)
the propagator for
the small component of the Dirac field of the heavy quark.  This last Feynman
rule is needed in
our calculation because the fragmenting heavy-quark is off its mass shell.
The Feynman rule for a heavy-quark propagator
is
\begin{equation}
\label{fey1}
\frac{i}{v\cdot k + \frac{C_1}{2m_Q} k^2 - \frac{C_2}{2m_Q}(v\cdot k)^2 } \;
\frac{1+\overlay{/}{v}}{2} \;,
\end{equation}
where $k$ is the residual 4-momentum of the heavy quark. The $QQg$ vertex
is
\begin{equation}
\label{fey2}
-ig_s T^a \, \left ( v^\mu + \frac{C_1}{2m_Q} (k_1+k_2)^\mu -
\frac{C_2}{2m_Q} v\cdot (k_1+k_2) v^\mu +
i \frac{C_3}{2m_Q} \sigma^{\mu\nu} q_\nu \right ) \;,
\end{equation}
where $k_1$ and $k_2$ are the residual 4-momenta of the incoming and outgoing
quarks and $q=k_2-k_1$ is the momentum of the gluon.
The Feynman rule for the propagator of the small component of the Dirac
field of the heavy-quark is
\begin{equation}
\frac{i}{v\cdot k}\,
\frac{1+\overlay{/}{v}}{2}\, \left( \frac{1}{2m_Q} \sigma^{\mu\nu} q_\nu
\right )\, \frac{1-\overlay{/}{v}}{2} \;.
\end{equation}

To calculate the fragmentation functions, we follow the method introduced in
Refs.~\cite{gluon} and \cite{psi} and applied in Ref.~\cite{bc} to the
fragmentation processes $\bar b\to B_c$ and $\bar b\to B_c^*$.
We denote the pseudoscalar and vector $Q\bar q$ mesons by $P$ and $V$,
respectively. Here $Q$ is the heavy quark, and $\bar{q}$ is the light
antiquark.    We calculate
the cross section for producing a $Q\bar q$ meson plus a light quark $q$ with
total 4-momentum $K^\mu$, divide it by the cross section for producing an
on-shell $Q$ with the same 3-momentum $\vec K$, and take the limit $K_0\to
\infty$. The fragmentation function is
\begin{equation}
\label{basic}
D(z) = \frac{1}{16\pi^2} \int ds \, \theta \left(s - \frac{M^2}{z} -
\frac{m_q^2}{1-z} \right ) \;\lim_{K_0\to \infty} \, \frac{\sum |{\cal M}|^2}
{\sum |{\cal M}_0|^2 } \;,
\end{equation}
where $M=m_Q+m_q$ is the mass of the meson in the nonrelativistic
approximation, $s=K^2$, ${\cal M}$ is the matrix element for producing
$P + q$ or $V + q$, and ${\cal M}_0$ is the matrix element for producing an
on-shell $Q$.   The calculation can be greatly simplified by using the axial
gauge with the gauge parameter $n^\mu=(1,0,0,-1)$ in the frame where
$K^\mu=(K_0,0,0,\sqrt{K_0^2-s})$.  In this gauge, we need only consider
the production of the $Q\bar q$ meson plus $q$ through a virtual $Q$
of momentum $K^\mu$.  The part of the matrix element ${\cal M}$ that involves
production of the virtual $Q$ can be treated as an unknown Dirac spinor
$\Gamma$.  In the limit $K_0\to \infty$, the same spinor factor $\Gamma$
appears in the matrix element ${\cal M}_0=\Gamma u(K)$  for an on-shell $Q$.
   The Feynman diagram for $Q^* \to Q\bar q + q$ is shown in
Fig.~\ref{fig2}.  The usual projection of the $Q\bar q$
onto a nonrelativistic $^1S_0$ bound state reduces in the heavy-quark
limit to the Feynman rule
\begin{equation}
Q\bar{q} \to {\delta^{ij}\over \sqrt{3}}  \; {R(0)\sqrt{M} \over \sqrt{4\pi}}
\;\gamma^5{1+\overlay{/}{v}\over 2},
\label{wf}
\end{equation}
where $R(0)$ is the radial wavefunction at the
origin for the meson and $v^\mu$ is its 4-velocity.
For the $^3S_1$ state, the projection is the same except
that $\gamma^5$ is replaced by $\overlay{/}{\epsilon}$, where
$\epsilon^\mu$ is the polarization 4-vector for the vector meson $V$.
The rest of the amplitude corresponding to Fig. \ref{fig2} is obtained by
using the ordinary QCD Feynman rules for the light quark spinor and the
$q\bar q g$  vertex and HQET Feynman rules for the heavy-quark propagator
and the $Q\bar{Q}g$ vertex.

The amplitude $\cal M$ for producing the $^1S_0$ state, including $1/m_Q$
corrections in the heavy-quark propagator and vertex, is
\begin{eqnarray}
\label{all}
\lefteqn{
i{\cal M} = - \frac{8\sqrt{\pi} \alpha_s R(0)}{3} \frac{M^2\sqrt{M}}{m_q}
\frac{1}{(s-m_Q^2)^2} \frac{1}{1+ \frac{C_1 m_q}{m_Q} - \frac{C_2}{2m_Q}v\cdot
k } \left( g_{\mu\nu} -\frac{n_\mu k_\nu +k_\mu n_\nu}{n\cdot k} \right )  }
\nonumber \\
&\times & \bar u(p') \gamma^\mu \gamma^5
(1+\overlay{/}{v} ) \left (
v^\nu + \frac{C_1}{2m_Q} k^\nu - \frac{C_2}{2 m_Q}(v\cdot k) v^\nu +
\frac{C_3}{4m_Q}( \gamma^\nu \overlay{/}{k} -\overlay{/}{k}\gamma^\nu ) \right
)\,\frac{1+\overlay{/}{v}}{2}\, \Gamma
\end{eqnarray}
where $k=m_q v + p'$ is the momentum of the virtual gluon and also the residual
momentum of the fragmenting heavy quark: $K=m_Q v + k$.
Note that the term proportional to $n_\nu$ in the numerator of the
axial-gauge propagator for the gluon vanishes after contracting with the Dirac
factor.   For the vector meson state,  the $\gamma^5$ in the above
equation is replaced by $\overlay{/}{\epsilon}$.

We are interested  only in the sum of the first two terms $a(y)/r + b(y)$
in the heavy-quark mass expansion, where $r=m_q/(m_Q+m_q)$.
We calculate separately the contributions to the fragmentation functions from
the leading terms in the HQET Feynman rules, from
the $1/m_Q$ corrections from the propagator,  and from
the $1/m_Q$ corrections from  the vertex.
For the following we will detail the derivation for the $^1S_0$ state,
but only quote the results for the $^3S_1$ state.

We first derive the fragmentation function $D_{Q\to P}(z)$
with the leading terms in the HQET propagator and vertex only.  The amplitude
reduces to
\begin{equation}
\label{11}
i{\cal M}_1 =  \frac{8\sqrt{\pi} \alpha_s R(0)}{3} \frac{M^2\sqrt{M}}{m_q}
\frac{1}{(s-m_Q^2)^2} \;
\bar u(p') \left( 1+  \frac{v\cdot k}{n\cdot k} \overlay{/}{n} \right )
\gamma^5 (1+\overlay{/}{v}) \, \Gamma \,.
\end{equation}
Squaring  and summing over spins and colors of the light  quark, we get
\begin{eqnarray}
\sum |{\cal M}_1|^2 &=& \frac{64\pi \alpha_s^2 |R(0)|^2}{9} \frac{M^5}{m_q^2}
 {\rm Tr}\left ( \Gamma \bar \Gamma (1+\overlay{/}{v}) \right )
\; \left [ \frac{z(1-z)}{M^3(1-(1-r)z)^2 (s- m_Q^2)^2}  \right. \nonumber \\
&& \left. \qquad + \frac{-1+z+3rz}{M(1-(1-r)z)(s-m_Q^2)^3}
- \frac{4rM}{(s-m_Q^2)^4} \right ] \;.
\end{eqnarray}
The corresponding amplitude-squared for producing an on-shell heavy quark is
\begin{equation}
\sum |{\cal M}_0|^2 = \frac{3M}{z} {\rm Tr}\left ( \Gamma \bar \Gamma
(1+\overlay{/}{v} )\right ) \;.
\end{equation}
Substituting $|{\cal M}|^2$ and $|{\cal M}_0|^2$ into (\ref{basic})
and integrating over $s$,  we get
\begin{eqnarray}
D_{Q\to P}(z) &=& \frac{2\alpha_s(2m_q)^2 |R(0)|^2}{81\pi m_q^3}
\frac{rz^3(1-z)^2}{(1-(1-r)z)^6} \nonumber \\
&&\times \left( 3(1-(1-r)z)^2 -8rz(1-z) +12rz (1-(1-r)z )  \right ) \;.
\end{eqnarray}
Expressing this in terms of $y$ using (\ref{no1})
and expanding to next-to-leading order in $r$,
we have
\begin{equation}
D_{Q\to P}(z) = N\frac{(y-1)^2}{ry^6} (3y^2+4y +8) - N\frac{(y-1)^3}{y^6}
(3y^2+4y+8 ) + {\cal O}(r)  \;,
\end{equation}
where $N=2\alpha_s^2|R(0)|^2/(81\pi m_q^3)$.
Therefore, in terms of $a(y)$ and $b(y)$, the leading term in the HQET
Lagrangian contributes
\begin{eqnarray}
a(y) &=& N\frac{(y-1)^2}{y^6} (3y^2+4y +8 ) \;, \\
\label{b1}
b_1(y) &=&  N\frac{(y-1)^2}{y^6} \left( -(y-1)( 3y^2 +4y +8 ) \right) \;.
\end{eqnarray}
The corresponding calculation for the $^3S_1$ state gives
$a^*(y)=3a(y)$ and $b^*_1(y)=3b_1(y)$.
These  contributions to  $D_{Q\to P}(z)$ and $D_{Q\to V}(z)$ differ
by a spin factor of 3, as required by  heavy-quark spin symmetry.

Next we calculate the contributions from $1/m_Q$ corrections in the
heavy-quark propagator and the heavy-quark vertex.  Expanding out
the $1/m_Q$ correction to the  propagator in (\ref{all}) to first order,
the correction to the amplitude is
\begin{eqnarray}
i{\cal M}_2 &=& \frac{8\sqrt{\pi} \alpha_s R(0)}{3} \frac{M^2\sqrt{M}}{m_q}
\frac{1}{(s-m_Q^2)^2} \left(- C_1 \frac{m_q}{m_Q} +\frac{C_2}{2m_Q}(m_q
+v\cdot p') \right) \nonumber \\
&& \times \; \bar u(p') \left( 1 + \frac{v\cdot k}{n\cdot k} \overlay{/}{n}
\right) \gamma^5 (1+\overlay{/}{v})  \,\Gamma \,.
\end{eqnarray}
Keeping the interference terms in $|{\cal M}_1 + {\cal M}_2|^2$,
summing over spins and colors, and inserting into (\ref{basic}), we
find a $1/m_Q$ correction  to $D_{Q\to P}(z)$.  Expressing this in terms
of $y$, we find that the contribution to $b(y)$ is
\begin{equation}
\label{b2}
b_2(y) =  N \frac{(y-1)^2}{y^6} \left( (-2C_1 + C_2 y )(3y^2+4y+8) \right )
\;.
\end{equation}
A similar calculation for the $^3S_1$ state gives  $b_2^*(y) = 3 b_2(y)$.
The $1/m_Q$ correction to the amplitude in (\ref{all}) from the heavy-quark
vertex is
\begin{eqnarray}
i{\cal M}_3 &=&- \frac{8\sqrt{\pi} \alpha_s |R(0)|}{3} \frac{M^2\sqrt{M}}{m_q}
\frac{1}{(s-m_Q^2)^2}
\left( g_{\mu\nu} -\frac{n_\mu k_\nu}{n\cdot k} \right ) \;
 \bar u(p') \gamma^\mu \gamma^5 (1+\overlay{/}{v} ) \nonumber \\
&& \times  \left (
 \frac{C_1}{2m_Q} k^\nu - \frac{C_2}{2 m_Q}(v\cdot k) v^\nu +
\frac{C_3}{4m_Q}( \gamma^\nu \overlay{/}{k} -\overlay{/}{k}\gamma^\nu ) \right
) \frac{1+\overlay{/}{v}}{2} \Gamma \;.
\end{eqnarray}
Keeping the interference terms in $|{\cal M}_1 + {\cal M}_3|^2$,  we obtain
after some work the contribution to $b(y)$ and to $b^*(y)$ due to the
$1/m_Q$ vertex correction
\begin{equation}
\label{b3}
b_3(y) = N\frac{(y-1)}{y^5} \left( -C_2 (y-1)(3y^2+4y+8)
+ 6C_1  (y-1) (y+2) - 12 C_3 y \right )  \;,
\end{equation}
\begin{equation}
\label{b3*}
b_3^*(y) =  3 N\frac{(y-1)}{y^5} \left( -C_2 (y-1)(3y^2+4y+8 )
+ 6 C_1 (y-1) (y+2) +  4 C_3 y \right ) \;.
\end{equation}

In (\ref{all}), the $(1+\overlay{/}{v})/2$ factor adjacent to $\Gamma$
projects onto the {\it large} component of the heavy-quark spinors produced
by the source $\Gamma$.
There is also a contribution of order
$1/m_Q$ from the {\it small} component of the heavy-quark spinors of the
fragmenting $Q$ quark \cite{jaffe}.  The corresponding amplitude is given by
\begin{eqnarray}
i{\cal M}_4 &=&- \frac{8\sqrt{\pi} \alpha_s R(0)}{3} \frac{M^2\sqrt{M}}{m_q}
\frac{1}{(s-m_Q^2)^2}
\left( g_{\mu\nu} -\frac{n_\mu k_\nu}{n\cdot k} \right ) \nonumber \\
&&\times \bar u(p') \gamma^\mu \gamma^5
(1+\overlay{/}{v} ) \left (
\frac{1}{4m_Q}( \gamma^\nu \overlay{/}{k} -\overlay{/}{k}\gamma^\nu ) \right
) \frac{(1 - \overlay{/}{v})}{2} \Gamma  \;.
\end{eqnarray}
The contributions  to $b(y)$ and to $b^*(y)$ from the interference term in
$|{\cal M}_1 + {\cal M}_4|^2$ are
\begin{equation}
\label{b4}
b_4(y) = 2N \frac{y-1}{y^5} (3y^3 +5y^2 +2y -4 )  \;,
\end{equation}
\begin{equation}
\label{b4*}
b_4^*(y) = 6N \frac{y-1}{y^5} ( y^3  - y^2 + 2y -4 )  \;.
\end{equation}

The complete expression for $b(y)$ is obtained by adding (\ref{b1}),
(\ref{b2}), (\ref{b3}), and (\ref{b4}).   Thus the fragmentation function
$D_{Q\to P}(z)$ for the $^1S_0$ state, to next-to-leading order in $1/m_Q$,
is given by (\ref{dz1}) with
\begin{eqnarray}
\label{ay}
a(y) &=& N\frac{(y-1)^2}{y^6} (3y^2 + 4y +8 )\;, \\
\label{by}
b(y) &=& N\frac{y-1}{y^6} \left( (y-1)(3y^3+15y^2+8y-8) -12 (C_3-1) y^2
\right ) \;.
\end{eqnarray}
The complete expression for $b^*(y)$ is obtained by
adding $3b_1(y)$, $3b_2(y)$, (\ref{b3*}), and (\ref{b4*}).
The fragmentation function $D_{Q\to V}(z)$ for the $^3S_1$ state,
to next-to-leading order in $1/m_Q$, is given by (\ref{dz2}) with
\begin{eqnarray}
\label{a*y}
a^*(y) &=& 3 N\frac{(y-1)^2}{y^6} (3y^2 + 4y +8 ) \;, \\
\label{b*y}
b^*(y) &=& 3 N\frac{y-1}{y^6} \left( -(y-1)(y^3 + y^2-8y+8) + 4 (C_3-1) y^2
\right ) \;.
\end{eqnarray}
The terms proportional to $C_2$  in (\ref{by}) and (\ref{b*y}) cancel
between propagator and vertex corrections.  We have set $C_1=1$ in (\ref{by})
and (\ref{b*y}).   If we further put $C_3=1$, we
recover the next-to-leading terms in the $1/r$ expansion of the PQCD
fragmentation functions given in Ref.~\cite{bc}.

The heavy-quark mass expansions (\ref{dz1}) and (\ref{dz2})  break down in the
limit $y\to\infty$, which corresponds to $z\to 0$, and also in the limit
$y\to 1$, which corresponds to $z\to 1$.  As $y\to \infty$, the leading terms,
given by (\ref{ay}) and (\ref{a*y}), scale like $1/(ry^2)$, while the
next-to-leading terms in (\ref{by}) and (\ref{b*y}) scale like $1/y$.  Thus
the $1/m_Q$ expansion breaks down when $y$ is of order $1/r$ or larger.
In  the limit
$y\to 1$, the leading terms in (\ref{dz1}) and (\ref{dz2}) vanish like
$(y-1)^2/r$ while the terms proportional to $C_3-1$ in the next-to-leading
terms go to 0 as the first power of $y-1$.  Thus, unless $C_3=1$, the
expansion also breaks down for $y-1$ of order $r$ or smaller.

In Fig.~\ref{fig3}, we compare the PQCD fragmentation functions (solid curves)
with the heavy-quark mass expansions (\ref{dz1}) and (\ref{dz2}) at leading
order (dotted curves) and next-to-leading order (dashed curves) in $r$.  We
use the value $r=0.10$, which corresponds to $D$ mesons. The normalization is
fixed by arbitrarily setting $N=1$ in (\ref{ay})-(\ref{b*y}). Note that we have
set $C_3=1$ in (\ref{by}) and (\ref{b*y}). For any other value of $C_3$, either
(\ref{dz1}) or (\ref{dz2}) becomes negative for $y$ very close to 1 indicating
the breakdown of the $1/m_Q$ expansion when $z$ is too close to 1. From the
figure it is clear that the next-to-leading order curves are in very good
agreement with the complete PQCD fragmentation functions for both $D$ and $D^*$
mesons. Surprisingly, the leading order result for fragmentation into $D^*$
mesons also agrees very well with the complete PQCD result, while the leading
order result for the $D$ meson falls about $30 \%$ low near the peak.

\begin{center}\section{PQCD  Model for Heavy-Quark Fragmentation}\end{center}

It is tempting to use the heavy-quark limits of the PQCD fragmentation
functions  as phenomenological models for the fragmentation of a heavy
quark $Q$ into heavy-light mesons $Q\bar  q$, where $Q=c$ or $b$ and
$q=u,d$, or $s$.
To next-to-leading order in $1/m_Q$, these fragmentation functions are given
by (\ref{dz1}) and (\ref{dz2}), with $a(y)$, $b(y)$, $a^*(y)$, and $b^*(y)$
given in (\ref{ay}) -- (\ref{b*y}).
In addition to $N$ and $r$, we must treat $C_3$ as a phenomenological
parameter, since, according to (\ref{c1-c3}), it depends on the low-energy
scale $\mu$ where perturbation theory breaks down.  These 3 parameters all
have well-defined scaling behavior with the heavy-quark mass; namely, $N$ is
independent of $m_Q$, $r$ scales like $1/m_Q$, and $C_3$ scales like
$\alpha_s(m_Q)^{9/(33-2n_f)}$.  Thus, if the parameters are determined
phenomenologically from data on charm
fragmentation into $D$ and $D^*$ mesons, then the corresponding parameters for
the $B$ and $B^*$ mesons can be determined by scaling.
The problem  with this model is that
unless $C_3=1$, either $D_{Q\to P}(z)$ or $D_{Q\to V}(z)$ becomes negative
for $z$ near 1.  This unphysical behavior only arises in a region of $z$ where
the $1/m_Q$ expansion is breaking down, but it makes these fragmentation
functions less attractive as a phenomenological model.  If we choose $C_3=1$
to avoid these difficulties, we might as well avoid the $1/m_Q$ expansion
altogether and use the complete PQCD fragmentation functions as our model.
We therefore propose as a model of heavy quark fragmentation
the PQCD fragmentation functions calculated in Ref.~\cite{bc}:
\begin{eqnarray}
\label{pqcd1}
D_{Q\to P}(z) &=&  N\, \frac{rz(1-z)^2}{(1-(1-r)z)^6}
\left [ 6 - 18(1-2r)z + (21 -74r+68r^2) z^2 \nonumber  \right. \\
&& \left. -2(1-r)(6-19r+18r^2)z^3  + 3(1-r)^2(1-2r+2r^2)z^4 \right ]\;, \\
\label{pqcd2}
D_{Q\to V}(z) &=& 3N \,\frac{rz(1-z)^2}{(1-(1-r)z)^6}
\left [ 2 - 2(3-2r)z + 3(3 - 2r+ 4r^2) z^2 \nonumber \right. \\
&& \left. -2(1-r)(4-r +2r^2)z^3  + (1-r)^2(3-2r+2r^2)z^4 \right ] \;.
\end{eqnarray}
The only parameters are the normalization $N$, which is independent of $m_Q$,
and $r$, which scales like $1/m_Q$.  The parameter $r$, which in the PQCD
calculation has the value $m_q/(m_Q+m_q)$, can be interpreted as the ratio
of the constituent mass of the light quark to the mass of the meson.
Integrating over $z$, we obtain the total fragmentation probabilities:
\begin{eqnarray}
\int_0^1 dz \; D_{Q\to P}  (z) &=& 3N\,\left(
\frac{8+13r+228r^2-212r^3+53r^4}{15(1-r)^5}  \right. \nonumber \\
&& \qquad \left.  + \frac{r(1+8r+r^2-6r^3+2r^4)\,\log(r)}{(1-r)^6} \right )
\label{32} \\
\int_0^1 dz \; D_{Q\to V}(z) &=& 3N\,\left(
\frac{24+109r-126r^2+174r^3+89r^4}{15(1-r)^5} \right. \nonumber \\
&& \qquad \left.  + \frac{r(7-4r+3r^2+10r^3+2r^4)\,\log(r)}{(1-r)^6} \right )
\label{33}
\end{eqnarray}

The PQCD fragmentation functions (\ref{pqcd1}) and (\ref{pqcd2}) give the
distributions in the longitudinal momentum fraction $z$ for the mesons $P$ and
$V$ in a heavy-quark jet.  This model can be easily extended to give the
distribution in their transverse momentum $k_T$ relative to the jet momentum
\cite{bc-pol}.
In Ref.~\cite{bc}, the fragmentation functions were obtained as integrals over
the invariant mass $s$ of the fragmenting heavy quark:
\begin{equation}
\label{dzs}
D_{Q\to P/V}(z) = \int_{s_{\rm min}(z)}^\infty
\frac{ds}{s} \; d_{Q\to P/V}(z,s) \;,
\end{equation}
where the lower limit of the integration is:
\begin{equation}
s_{\rm min}(z) = { M^2 \over z} + {r^2 M^2 \over 1-z}.
\label{smin}
\end{equation}
The functions $d_{Q \to P/V}(z,s)$ in the integrand are given by
\begin{eqnarray}
d_{Q \to P} (z,s) &=& 6 N M^2 r s \left[
\frac{(1-z)(1+rz)^2 }{(1-(1-r)z)^2(s-(1-r)^2 M^2)^2}
\right. \nonumber \\
& - &  \left.
 \frac{[2(1-2r) -(3-4r+4r^2)z +(1-r)(1-2r)z^2] M^2}{(1-(1-r)z)
(s-(1-r)^2 M^2)^3}
- \frac{4r(1-r) M^4}{(s-(1-r)^2 M^2)^4}
\right ] \nonumber \\
&& \; \\
d_{Q \to V} (z,s) &=& 6 N M^2 r s \left[
\frac{(1-z)(1+2rz+(2+r^2)z^2) }{(1-(1-r)z)^2(s-(1-r)^2 M^2)^2}
\right. \nonumber \\
& - &  \left.
 \frac{[2(1+2r) -(1+12r-4r^2)z - (1-r)(1+2r)z^2] M^2}{(1-(1-r)z)
(s-(1-r)^2 M^2)^3}
- \frac{12r(1-r) M^4}{(s-(1-r)^2 M^2)^4}
\right ] \nonumber\\
&& \;
\end{eqnarray}
The invariant mass $s$ is related to $k_T$ and $z$ by
\begin{equation}
\label{stokt}
s = \frac{M^2+k_T^2}{z} + \frac{m_q^2 + k_T^2}{1-z}\;,
\end{equation}
where $M=m_Q+m_q$ in the nonrelativistic limit.  If, instead of integrating
over $s$, we integrate over $z$ with $k_T^2$ held fixed, we obtained the $k_T$
distribution for the fragmentation process.
Introducing the dimensionless variable $t=k_T/M$, we can define the
$k_T$-dependent functions $d_{Q\to P/V}(z,t)$ and $D_{Q\to P/V}(t)$ by
\begin{eqnarray}
\int_0^\infty dt \;D_{Q\to P/V}(t) &=& \int_0^\infty dt \int_0^1 dz \; d_{Q\to
P/V}(z,t) \\
&=& \int_0^1 dz \int_{s_{\rm min}(z)}^\infty \frac{ds}{s} \;
d_{Q\to P/V}(z,s) \;.
\end{eqnarray}
This implies
\begin{equation}
D_{Q\to P/V}(t) = 2M^2 t \int_0^1 dz {1 \over z(1-z)s(z,t)}
\;d_{Q\to P/V}(z,s(z,t))
\end{equation}
with
\[
s(z,t)=M^2\left( \frac{1+t^2}{z} + \frac{r^2+t^2}{1-z} \right)
\]
Carrying out the integrals over $z$, we find
\begin{eqnarray}
D_{Q\to P}(t) &=& \frac{Nr}{2(1-r)^6}\,\frac{1}{t^6}
\Bigg \{
-24 r t \left[ 4 r^2 - (2 + r + 2 r^2)t^2 \right] \log(r)   \nonumber \\
&-& (1-r)t \left[ 30r^3 -r(61-20r+28r^2)t^2 -(3-48r +48r^2 -12r^3)t^4
\right ] \nonumber \\
&+& 12t \left[ 4r^3-r(2+r+2r^2)t^2 +(1-r)^2 t^6\right]
\log \Biggl( \frac{r^2+t^2}{1+t^2} \Biggr) \nonumber \\
&+& 3 \left[ 10r^4 -3r^2(11+2r+2r^2)t^2 +(3+4r+19r^2-6r^3)t^4
        \right .\nonumber \\
&& \left. + (3+12r-20r^2+8r^3)t^6 \right ]\,
{\rm Arctan}\left( \frac{(1-r)t}{r + t^2} \right ) \Bigg \}
\label{44}
\end{eqnarray}
\begin{eqnarray}
D_{Q\to V}(t) &=& \frac{3Nr}{2(1-r)^6}\,\frac{1}{t^6}
\Bigg \{
- 8 r t \left[ 12 r^2 - (6 + 7 r + 2 r^2) t^2 \right]  \log(r)   \nonumber \\
&-& (1-r)t \left[ 30r^3 -r(61+28r-20r^2)t^2 +(5-8r +8r^2 +4r^3)t^4
\right ] \nonumber \\
&+& 4t \left[ 12r^3-r(6+7r+2r^2)t^2 +(1-r)^2 t^6\right]
\log \Biggl( \frac{r^2+t^2}{1+t^2} \Biggr) \nonumber \\
&+& \left[ 30r^4 - 3r^2(33+22r-10r^2)t^2 +(9+20r+r^2+22r^3+8r^4)t^4
        \right .\nonumber \\
&& \left. + (9-12r+4r^2+8r^3)t^6 \right ]\, {\rm Arctan}\left( \frac{(1-r)t}{r
+ t^2} \right ) \Bigg \} \;.
\label{45}
\end{eqnarray}

 In general, fragmentation functions $D(z,\mu^2)$ depend not only on $z$ but
also on a factorization scale $\mu$.  In a high energy process that produces
a jet with transverse momentum $p_T$, the scale $\mu$ should be chosen to be
on the order of $p_T$.  The functions (\ref{pqcd1}) and (\ref{pqcd2}) should
be regarded as models for heavy-quark fragmentation functions at a scale $\mu$
of order $m_Q$.  For values of $\mu$ much larger than $m_Q$, the fragmentation
functions (\ref{pqcd1}) and (\ref{pqcd2}) should be evolved from the scale
$m_Q$ to the scale $\mu$ using the Altarelli-Parisi equation:
\begin{equation}
\label{evol}
\mu^2 \frac{\partial}{\partial \mu^2}\, D_{Q\to H}(z,\mu^2) = \int_z^1
\frac{dy}{y} \; P_{Q\to Q}(\frac{z}{y},\mu)\, D_{Q\to H}(y,\mu^2) \;,
\end{equation}
where $P_{Q\to Q}(x)$ is the appropriate splitting function:
\begin{equation}
P_{Q\to Q}(x,\mu) = \frac{2\alpha_s(\mu)}{3\pi} \left( \frac{1+x^2}{1-x}
\right)_+ \;.
\end{equation}

One aspect of the initial conditions (\ref{pqcd1}) and (\ref{pqcd2}) and the
evolution equation (\ref{evol}) that may cause problems in practical
applications is that they do not respect the phase space constraint:
\begin{equation}
\label{const}
D_{Q\to H}(z,\mu^2) = 0 \quad {\rm for} \quad z < M^2/\mu^2 \;,
\end{equation}
This can be remedied  \cite{BDFM} by using (\ref{const}) as the initial
condition on the fragmentation function
equation and replacing (\ref{evol}) by the inhomogeneous evolution equation
\begin{equation}
\mu^2 \frac{\partial}{\partial \mu^2}\, D_{Q\to H}(z,\mu^2) = \int_z^1
\frac{dy}{y} \; P_{Q\to Q}(\frac{z}{y},\mu)\, D_{Q\to H}(y,y\mu^2)
+ d_{Q\to H}(z,\mu^2) \theta(\mu^2 - s_{\rm min}(z) ) \;,
\end{equation}
where $d_{Q\to H}(z,s)$ is defined by the integrand in (\ref{dzs}) and
$s_{\rm min}(z)$ is given in (\ref{smin}).

\begin{center}\section{Comparison with other Fragmentation Models}
\end{center}

The model for heavy-quark fragmentation which has been used most extensively
in phenomenological applications is the Peterson fragmentation function
\cite{peterson}:
\begin{equation}
\label{eq-peter}
D_{Q\to H}(z) = N_H\; \frac{z(1-z)^2} { [(1-z)^2 + \epsilon_H z ]^2 }\;,
\end{equation}
where $N_H$ and $\epsilon_H$ are adjustable parameters that may depend on the
hadron $H$.  This fragmentation function has the correct behavior
in the heavy-quark limit if $N_H$ scales like $1/m_Q$ and $\epsilon_H$
scales like $1/m_Q^2$.  Identifying $\epsilon_H$ with $r^2$ and expressing
(\ref{eq-peter}) in terms of the Jaffe-Randall scaling variable $y$ defined in
(\ref{no1}), we find that it reduces in the limit $r\to 0$ to
\begin{equation}
D_{Q\to H}(z)  \to \frac{N_H}{r^2} \frac{(y-1)^2}{[(y-1)^2+1]^2} \;.
\end{equation}
The Peterson fragmentation function is just the square of a light-cone energy
denominator multiplied by a phase space factor.  It
contains no spin information; the normalization parameter $N_H$
is to be determined independently for the pseudoscalar and vector mesons of a
heavy-quark spin multiplet.

An alternative fragmentation model which does contain spin information
has been proposed by Suzuki \cite{suzuki}.  Suzuki's fragmentation functions
are derived from the same Feynman diagram in Fig.~\ref{fig2} as the PQCD
fragmentation functions,   but with two essential differences.
First, the diagram was calculated in Feynman gauge.  If a general covariant
gauge had been used, Suzuki's fragmentation functions would have depended
on the gauge parameter.  The PQCD fragmentation functions that we calculated
are gauge-invariant.
We
calculated the diagram in the axial gauge only for simplicity.  If we had used
a covariant gauge, we would have had to also include diagrams in which both
the virtual heavy quark and the virtual gluon are emitted by the source
$\Gamma$
in Fig.~\ref{fig2}.  Alternatively, we could have calculated the PQCD
fragmentation functions for the fragmentation of a heavy quark into S-wave
heavy
quarkonium directly from the general gauge-invariant definition
\cite{collin}. Such a calculation has been carried out for the equal mass case
of charmonium by Ma \cite{ma}.
A second essential difference between the PQCD model and Suzuki's is that we
integrated over the invariant mass  $s$ of the fragmenting quark (see
Eq.~(\ref{basic})).  The invariant mass is related to the transverse momentum
$k_T$ of the meson relative to the fragmenting quark by (\ref{stokt}).
%
Rather than
integrating over $k_T^2$, Suzuki chose to evaluate the integrand at a typical
value $\langle k_T^2 \rangle$.
Suzuki's model therefore has 3 parameters: the overall
normalization $N$, the mass ratio $r$, and $\langle k_T^2 \rangle /m_Q^2$.
When expressed in terms of the scaling variable $y$ defined in (\ref{no1}),
Suzuki's fragmentation function $D_{Q\to P}(z)$ reduces in the limit $r\to 0$
to
\begin{equation}
\label{suzlim}
D_{Q\to P}(z) \to \frac{N}{r} \, (y-1)^2 \, \frac{(y-2)^2 + \kappa^2 }
{[y^2 + \kappa^2 ]^4} \;,
\end{equation}
where $\kappa^2 = \langle k_T^2 \rangle /(r^2m_Q^2)$. By heavy-quark spin
symmetry $D_{Q\to V}(z)$ differs, in this limit, only by a factor of 3.

The Peterson, Suzuki, and PQCD fragmentation functions all vanish like
$(1-z)^2$ as $z\to 1$.
An alternative  fragmentation  function which vanishes like the first power of
$(1-z)$  has been proposed by Collins and Spillers
\cite{collins}. This was motivated by incorrect dimensional counting
rules. The correct dimensional counting rules for QCD \cite{cr} do in fact give
a limiting behavior of $(1-z)^2$ for the fragmentation function. The
Collins-Spillers fragmentation function can be derived in a similar way
to ours, except that in the Feynman diagram in Fig.~\ref{fig2}, the
interaction mediated by the virtual gluon is replaced by a point-like  scalar
Yukawa coupling between the meson, the heavy quark, and the light quark.
Consequently,  the denominator of the matrix element contains only one power
of $(s-m_Q^2)$, in contrast to the 2 powers in (\ref{11}).  It is the omission
of the gluon propagator that changes the behavior as $z\to 1$ from
$(1-z)^2$ to $(1-z)$.  Also, instead of integrating over the invariant mass of
the fragmenting quark as in (\ref{basic}), Collins and Spillers, like Suzuki,
evaluated the integral at a typical value $\langle k_T^2 \rangle$.
Taking the scaling limit $r\to
0$, the fragmentation function of Collins and Spillers reduces to
\begin{equation}
D_{Q\to P}(z) \to \frac{2N}{r}\, (y-1) \,
\frac{(y-2)^2 + \kappa^2}
     { [y^2 + \kappa^2]^2 } \;,
\end{equation}
where $\kappa^2 = \langle k_T^2 \rangle /(r^2 m_Q^2)$.

The various fragmentation models in the literature have been summarized in
Ref.~\cite{cleo} and compared with experimental data on $D$ and $D^*$
production.  The string models and parton cluster models are very different
in spirit from those discussed above.  One can derive analytic expressions
for the  heavy-quark fragmentation functions from the string models
\cite{lund}.  They contain a tunneling factor $\exp(-Bm_H^2/z)$, which
suppresses the small-$z$ region.  In the scaling limit, the Lund symmetric
fragmentation function behaves like
\begin{equation}
D_{Q\to H}(z) \to N r^\beta e^{-B(m_H^2 + \langle k_T^2 \rangle )} (y-1)^\beta
\;.
\end{equation}
Unless $N$ scales like $e^{B m^2_Q} m^{\beta+1}_Q$, this is inconsistent
with heavy-quark symmetry, which requires the leading
term to scale like $m_Q$ as $m_Q \to \infty$.

The PQCD model for heavy-quark fragmentation has a number of advantages over
those described above.  First, it is rigorously correct in the limit $m_q \gg
\Lambda_{\rm QCD}$.  Higher order perturbative corrections can be
systematically calculated.  Relativistic corrections can also be calculated in
terms of additional nonperturbative matrix elements \cite{BBL}.
Second,  our model is consistent with heavy-quark symmetry in the limit
$m_Q\to \infty$.  The logarithms of $m_Q$ that are predicted by HQET would be
reproduced by the higher order perturbative corrections.   The PQCD model is
also more predictive than those in Refs.~\cite{peterson,suzuki,collins}.
It describes spin-dependent effects, like Suzuki's model, but without
introducing any additional parameters.  The PQCD model not only  predicts the
$z$-dependence of the fragmentation functions but also their dependence on
$k_T$, the transverse momentum of the meson relative to the jet.   The
fragmentation functions (\ref{pqcd1}) and (\ref{pqcd2})  apply only to S-wave
mesons, but the fragmentation functions for higher orbital-angular-momentum
states can also be calculated.  The PQCD fragmentation functions for the
P-wave mesons have been calculated  to leading order in $\alpha_s$ in
Refs.~\cite{TC}.

\begin{center}\section{The Vector-to-Pseudoscalar Ratio}
\end{center}

In any production process for heavy-light mesons, one of the most fundamental
experimental observables is the ratio
\begin{equation}
\label{pv}
P_{V}=\frac{V}{V+P} \;,
\end{equation}
which measures the relative number of vector mesons $V$ and pseudoscalar
mesons $P$ that are produced.   If the mesons are produced within a
heavy-quark jet, then  $V$ and $P$ in (\ref{pv}) can be identified as the
fragmentation probabilities for the heavy quark to fragment  into vector and
pseudoscalar mesons, respectively.
The ratio $P_V$ can depend on kinematic variables, such as the longitudinal
momentum fraction $z$ of the meson or its transverse momentum $k_T$
relative to the jet.  In the PQCD model for fragmentation,  the normalization
factor $N$ cancels out in the ratio (\ref{pv}), so that $P_V$ is determined by
the parameter $r$ only.

The simplest measure of the ratio $P_V$ comes from
the total numbers of vector and pseudoscalar mesons in the jet integrated
over $z$ and $k_T$.  Setting $P$ and $V$ in (\ref{pv}) to the fragmentation
probabilities in (\ref{32}) and (\ref{33}), we find that
the ratio $P_V$ in the PQCD model of fragmentation is
\begin{equation}
P_V= \frac{(1-r)(24+109r-126r^2+174r^3+89r^4)
+ 15r(7-4r+3r^2+10r^3+2r^4)\log (r)}
{2(1-r)(16+61r+51r^2-19r^3+71r^4)+60r(2+r+r^2+r^3+r^4)\log (r)}
\end{equation}
This ratio  is plotted as a function of $r$ in Fig.~\ref{fig-pv}. From the
graph
it is clear that $P_V$ is not strongly dependent on r.  At $r=0$, $P_V=3/4$ as
required by heavy-quark spin symmetry. As $r$ increases $P_V$ decreases slowly
to $P_V = 0.51$ at $r=0.5$. Thus at nonzero values of $r$, the vector
state is less populated than would be given by naive spin counting.
We can determine the value of $r$ for the $D$ and $D^*$ system using
experimental measurements of $P_{V}$.  A complete compilation of experimental
data for $P_{V}$ from LEP, CLEO, ARGUS, PETRA, and  TRISTAN can be found in
Ref.~\cite{peskin}.
The key point in obtaining consistency between
these measurements is using the updated branching
ratio B$(D^{+*}\to D^0 \pi^+)\approx 0.68$ instead
of the old value 0.55.    The experimental value $P_V=0.65\pm0.06$ determines
the parameter $r_D$ for the $D-D^*$ system to be
$r_D=0.10 \plusm{0.12}{0.07}$.  If we
interpret $r$ as the ratio of the constituent mass of the light quark to the
mass of the meson, then the value $r_D=0.10$ corresponds to a constituent mass
of 200 MeV.  Given a value of $r_D$, we can determine the corresponding value
for the $B-B^*$ system by using the simple scaling behavior $r_B=(m_D/m_B)
r_D$.  This gives $r_B=0.03 \plusm{0.04}{0.02}$.

Having  determined the parameter $r$ from data on $D-D^*$ production,
we can now predict how the vector-to-pseudoscalar ratio should
vary as a function of the longitudinal momentum fraction $z$.  The
$z$-dependent ratio $P_V(z)$ is defined by (\ref{pv}), where $P$ and $V$ are
given by the fragmentation functions (\ref{pqcd1}) and (\ref{pqcd2}):
\begin{equation}
\label{pvz}
P_V(z)  = \frac{3}{4} \frac{n(z)}{d(z)} \; ,
\end{equation}
with
\begin{eqnarray}
 n(z) & = & 2 - 2(3-2r)z + 3 (3-2r+4r^2)z^2 \nonumber \\
&& - 2(1-r)(4-r+2r^2)z^3 + (1-r)^2(3-2r+2r^2)z^4  \; , \\
 d(z) & = & 3 - 3(3-4r)z + (12-23r+26r^2)z^2 \nonumber \\
&& -(1-r)(9-11r+12r^2)z^3 + 3(1-r)^2(1-r+r^2)z^4  \; .
\end{eqnarray}
This ratio is plotted as a function of $z$ in Fig.~\ref{fig5} for the values
$r=0.10$ (solid curve), $r=0.03$ (dotted curve), and $r=0.22$
(dashed curve).  At $z=0$, $P_V(z)=1/2$, regardless of the value of $r$. It
decreases slightly for small $z$, and then increases monotonically to a maximum
value, at $z=1$, of 0.74 for $r=0.03$, 0.73 for $r=0.10$, and 0.70 for
$r=0.22$.
Note that, in spite of the large uncertainty in our determination of $r$, the
uncertainty in $P_V(z)$ is less than about $11 \%$. Thus the PQCD model gives a
rather unambiguous prediction that $P_V$ should vary from around 1/2 at small
values of $z$ to almost 3/4 near $z=1$.

The $k_T$-dependent ratio $P_V(k_T)$ is defined by (\ref{pv}), with $P$ and
$V$ given by (\ref{44}) and (\ref{45}):
\begin{equation}
P_V(k_T)  =  \frac{3}{4} \; \frac{  n_1 + n_2 \log (r) + n_3 \log
			\left( \frac{r^2+t^2}{1+t^2} \right)
		+ n_4 {\rm Arctan} \left( \frac{(1-r)t}{r+t^2} \right)  }
	{ d_1 + d_2 \log (r) + d_3 \log
			\left( \frac{r^2+t^2}{1+t^2} \right)
		+ d_4 {\rm Arctan} \left( \frac{(1-r)t}{r+t^2} \right)
		}  \; ,
\end{equation}
where
\begin{eqnarray}
n_1 & = & - (1-r)t [30 r^3 - r(61+28r-20r^2)t^2+(5-8r+8r^2+4r^3)t^4] \; , \\
n_2 & = & - 8 r t [12 r^2 - (6+7r+2r^2)t^2]  \; , \\
n_3 & = & 4t[12r^3-r(6+7r+2r^2)t^2+(1-r)^2t^6] \; , \\
n_4 & = & [30r^4-3r^2(33+22r-10r^2)t^2+(9+20r+r^2+22r^3+8r^4)t^4 \nonumber \\
&&  \;\;\;\;\;\;\;\;		+(9-12r+4r^2+8r^3)t^6] \; ,
\end{eqnarray}
and
\begin{eqnarray}
d_1 & = & - (1-r)t [30 r^3 - r(61+16r-8r^2)t^2+3(1+2r-2r^2+2r^3)t^4] \; , \\
d_2 & = & - 12 r t [ 8r^2 - 2 (2+2r+r^2)t^2 ] \; , \\
d_3 & = & 6t[8r^3-2r(2+2r+r^2)t^2+(1-r)^2t^6] \; , \\
d_4 & = & 30r^4 - 9r^2(11+6r-2r^2)t^2+  3(1+r)^2 (3+2r^2) t^4
		+ 3(3-4r^2+4r^3)t^6 \; . \nonumber \\
&& \;
\end{eqnarray}
This ratio is plotted as a function of $t=k_T/M$ in Fig.~\ref{fig6} for the
three values $r=0.10, 0.03$, and 0.22.  At $t = 0$, $P_V(t)=3/4$, regardless of
the value of $r$. As $t$ increases, $P_V(t)$ quickly decreases to its
asymptotic
value at $t = \infty$. At $t=1$, $P_V(t)$ is within $0.1\%$ of its asymptotic
value of $0.65$ for $r=0.03$, $0.62$ for $r=0.10$, and $0.60$ for $r=0.22$.
Again we find that, in spite of the large uncertainty in $r$, we obtain a
rather
precise prediction for $P_V$ as a function of $k_T$.

The PQCD fragmentation functions for vector mesons have been applied previously
\cite{bc-pol} as a phenomenological model to describe the fragmentation
processes $c\to D^*$ and
$b\to B^*$.   The fragmentation functions were separated into the transverse
and longitudinal polarization components.
The spin alignment, which measures the ratio of transverse to  longitudinal
polarizations,  was calculated as a function of $z$ and as a function of
$k_T$.  In the case of production of $D^*$ by charm fragmentation,
the spin alignment  predicted by the PQCD fragmentation model
was shown to be consistent with CLEO measurements \cite{bc-pol}.
In addition, the predicted value of the average longitudinal momentum fraction
$\langle z \rangle$
for $c\to D^*$ and for $b\to B^*$ was shown to be in excellent agreement
with data from LEP, CLEO, and ARGUS \cite{bc-pol}.
The values of $r$ used for $D^*$ and $B^*$ mesons in these comparisons were
 $r=0.17$ and $r=0.058$, respectively,  which lie
within the range determined above from measurements of $P_V$.

The PQCD fragmentation functions have also been applied in Ref.~\cite{CO} to
predict the fragmentation spectra for the $B_s$ and $B_s^*$ mesons based on
the production rates of the $B_s$ mesons measured  at LEP.   Instead of
treating the normalization $N$ as a phenomenological parameter as advocated
in this paper, the authors calculated $N$ using the PQCD expression, which
involves $\alpha_s$ at the scale of the strange quark mass.

\begin{center}\section{Summary} \end{center}

We have studied the heavy-quark mass limit of the PQCD fragmentation functions
for producing S-wave mesons.  The leading and next-to-leading terms in $1/m_Q$
were calculated directly from HQET.  The PQCD fragmentation functions were
proposed as a phenomenological model for fragmentation into heavy-light
mesons.  With only 2 parameters, this model describes fragmentation into the
$^1S_0$ pseudoscalar meson state and the transverse and longitudinal
polarization states of the $^3S_1$ vector meson.   It describes not only the
$z$-dependence of the fragmentation probabilities, but also their dependence
on the transverse momentum $k_T$ of the meson relative to the jet within which
it is produced.
This model can easily be extended to describe heavy quark fragmentation
into $P$-wave states using the PQCD
fragmentation functions calculated in \cite{TC}.
The PQCD fragmentation functions were compared with other
models for heavy-quark fragmentation in the literature.  As an application,
the  PQCD fragmentation functions were used to predict the ratio of
vector-to-pseudoscalar states as a function of $z$ and as a function of $k_T$.
The ratio $P_V$ is predicted to vary from around 1/2 at small values of $z$ to
almost 3/4 near $z=1$.

\bigskip

\section*{Acknowledgements}

We are grateful to Michael Peskin and Mark Wise for discussions.
This work was supported by the U.~S. Department of Energy, Division of
High Energy Physics, under Grants DE-FG02-91-ER40684 and DE-FG03-91ER40674.
One of us (E.B) would like to express his gratitude to the Theory Group at
Fermilab for their hospitality while this research was being carried out.


\newpage
\begin{center}\section*{Figure Captions}\end{center}

\begin{enumerate}


\item
\label{fig2}
Feynman diagram used to calculate the PQCD fragmentation functions in axial
gauge.


\item
\label{fig3}
Comparison of the $D_{c\to D}(z)$ (lower set of curves) and
$D_{c\to D^*}(z)$ (upper set of curves) fragmentation functions. The
normalization is arbitrary. Shown are the full PQCD results (solid curves),
the leading terms (dotted curves) in the heavy-quark mass expansion, and the
leading plus next-to-leading terms (dashed curves) in the heavy-quark mass
expansion.

\item
\label{fig-pv}
The ratio $P_{V}$ as a function of $r$.

\item
\label{fig5}
Predictions for the ratio $P_V(z)$ as a function of $z$ for
$r=0.10$ (solid curve), $r=0.03$ (dotted curve), and $r=0.22$ (dashed curve).

\item
\label{fig6}
Predictions for the ratio $P_V(k_T)$ as a function of $k_T$ for
$r=0.10$ (solid curve), $r=0.03$ (dotted curve), and $r=0.22$ (dashed curve).

\end{enumerate}

\end{document}